\documentclass[screen,acmsmall,table]{acmart}

\usepackage{libertine}
\usepackage{xcolor}
\usepackage{balance}
\usepackage{color}
\usepackage{booktabs} 
\usepackage{url}
\usepackage{graphicx}
\usepackage{subfigure}
\usepackage{multirow}
\usepackage{xspace}
\usepackage{enumitem} 
\usepackage[normalem]{ulem}
\usepackage{soul}
\usepackage{balance}
\usepackage{comment}
\usepackage{amsmath}
\usepackage{microtype}
\usepackage{geometry}
\usepackage{tcolorbox}
\usepackage[frozencache,cachedir=minted-cache]{minted}

\setcopyright{none}
\begin{document}

\title{Beyond Accuracy: An Empirical Study on Unit Testing in Open-source Deep Learning Projects}

\author{Han Wang}
\email{han.wang@monash.edu}
\orcid{https://orcid.org/0000-0001-7862-6677}
\affiliation{%
  \institution{Faculty   of   Information   Technology,   Monash   University}
  \country{Australia}
}

\author{Sijia Yu}
\email{yusj19@mails.jlu.edu.cn}
\orcid{https://orcid.org/0009-0003-1946-9407}
\affiliation{%
  \institution{Jilin University}
  \country{Changchun, China}
}
\author{Chunyang Chen}
\email{ chunyang.chen@monash.edu}
\orcid{https://orcid.org/0000-0003-2011-9618}
\affiliation{%
  \institution{Faculty   of   Information   Technology,   Monash   University}
  \country{Australia}
}

\author{Burak Turhan}
\email{burak.turhan@oulu.fi}
\orcid{https://orcid.org/0000-0003-1511-2163}
\affiliation{%
  \institution{M3S Research Unit, Faculty of ITEE, University of Oulu}
  \country{Finland}
}
\authornote{Burak Turhan is the corresponding author.}

\author{Xiaodong Zhu}
\email{zhuxd@jlu.edu.cn}
\orcid{https://orcid.org/0000-0002-7200-7629}
\affiliation{%
  \institution{Jilin University}
  \country{Changchun, China}
}

\begin{abstract}
Deep Learning (DL) models have rapidly advanced, focusing on achieving high performance through testing model accuracy and robustness.
However, it is unclear whether DL projects, as software systems, are tested thoroughly or functionally correct when there is a need to treat and test them like other software systems. 
Therefore, we empirically study the unit tests in open-source DL projects, analyzing 9,129 projects from GitHub. We find that: 1) unit tested DL projects have positive correlation with the open-source project metrics and have a higher acceptance rate of pull requests, 2) 68\% of the sampled DL projects are not unit tested at all, 3) the layer and utilities (utils) of DL models have the most unit tests. Based on these findings and previous research outcomes, we built a mapping taxonomy between unit tests and faults in DL projects. We discuss the implications of our findings for developers and researchers and highlight the need for unit testing in open-source DL projects to ensure their reliability and stability. The study contributes to this community by raising awareness of the importance of unit testing in DL projects and encouraging further research in this area. 
	
\end{abstract}


\maketitle

\section{Introduction}
\label{sec:introduction}
In recent years, Deep Learning (DL)~\cite{lecun2015deep}, a key branch of Machine Learning (ML), has demonstrated significant performance in different Artificial Intelligence (AI) tasks such as autonomous driving~\cite{gupta2021deep}, medical image diagnosis~\cite{devunooru2021deep}, and speech recognition~\cite{park2019specaugment}. The open-source versions of successful DL projects became readily available on code-sharing websites, and people adopted them for use in their research and applications~\cite{shatnawi2018comparative}. As open-source projects are freely accessible, they have the potential to be used and incorporated into various downstream systems, leading to widespread adoption and usage. However, without proper testing, any issues within the projects may be propagated downstream, potentially causing problems from incorrectly interpreting a voice to text to a severe car crash in autonomous driving.

There have been many studies on model evaluation, i.e., evaluation of the performance (e.g., accuracy, precision, etc.) on a validation or test dataset~\cite{ma2019deepct,sekhon2019towards}.
To ensure consistent model performance in various environments, many approaches have been proposed to defend against potential adversarial attacks~\cite{ma2019deepct,pei2017deepxplore,tian2018deeptest, yan2019artdl}.
However, open-source DL models can be employed as integral components in software systems developed by others or incorporated into various research projects. As such, it is essential to conduct unit testing on these individual components to identify and mitigate potential faults within them, thereby ensuring that the model functions according to expectations when integrated into a larger system.

Previous research has summarized different types of faults from existing DL projects~\cite{humbatova2020taxonomy, islam2020repairing}.
Faults in a DL model may be due to the misconfiguration of learning parameters (seen in StackOverflow(SO) \href{https://stackoverflow.com/questions/49226447}{\#49226447}), receiving data in the wrong format for the tensor input (SO \href{https://stackoverflow.com/questions/41563720}{\#41563720}), etc. Sometimes even simply making API calls from Tensorflow can cause bugs (SO \href{https://stackoverflow.com/questions/49742061}{\#49742061}). 
In addition, specific bugs in DL models are deeply embedded in the code, e.g., the project will not generate any error messages or crashes if the network is not stacked, but will cause the model to underperform, resulting in extra effort to debug~\cite{web:whyunittest}.

Unit tests are widely used in conventional software systems to ensure software quality. They can be automated, serve as a source of documentation, and can detect problems at early stages before deployment~\cite{trautsch2017there, thummalapenta2011retrofitting,daka2014survey}. 
Nevertheless, unlike conventional software development with relatively clear logic, DL systems are fuzzier, with inherent randomness, massive datasets, and limited interpretability. 
Moreover, during our preliminary observations on the unit tests in DL projects, we noticed that the way DL projects are unit tested differs from conventional unit testing, e.g., new assertion statements for DL projects only, asserting the range or shape instead of an absolute value. 
Even within the same DL project, we found unit tests to be applied in different parts of a DL system have other preferences in testing properties, for example, pre-processing (e.g., checking the shape and scale of data), model architecture (e.g., matching the shape of the layer output), training (e.g., checking weight changes during training), and evaluation (e.g., checking the logged outputs are as expected). 
While prior research~\cite{nejadgholi2019study, jia2021unit} has delved into unit testing of DL libraries, providing valuable perspectives such as the use of oracle approximation assertions to validate output ranges~\cite{nejadgholi2019study}, there remains a gap in the comprehensive understanding of the motivations and methodologies of unit testing in DL projects.

To fill the knowledge gap, we conduct a large-scale empirical study on unit testing in open-source DL projects.
We raise the following three research questions with motivations:

\textbf{RQ1: How can unit tests help open-source DL projects?}
In conventional projects, unit testing is essential in indicating code quality and helps to debug, especially when making new changes to the code base~\cite{grano2020pizza}.
However, DL models are complex and highly dynamic, and it is unclear how unit tests can be applied to these projects or whether they are necessary. In this RQ, we examine various open-source project metrics and whether they have unit tests to gain insights into the association of unit tests with the popularity of open-source DL projects and to determine the importance of unit tests in this rapidly growing field.

\textbf{RQ2: To what extent are current DL projects unit tested?}

As stated, the trend of using open-source DL models is rising. However, with thousands of models introduced each year, are developers writing unit test cases for the newly introduced models; if so, which testing frameworks are they using? The findings answering this research question shall provide a general picture of unit testing in the open-source DL field.

\textbf{RQ3: Which units and properties are tested in DL projects?}

For DL models that already have unit test cases, we aim to understand what parts of a DL model and which property of the unit the developers are most concerned with. Results shall help us uncover testing gaps, identify potential problem areas, and improve future development efforts in unit testing for open-source DL projects.

RQ1 seeks to establish the foundational understanding of the significance of unit tests in DL projects. Building upon this understanding, RQ2 narrows down to the actual adoption rates. It quantifies the prevalence of unit tests in the DL domain, providing a quantitative perspective on how extensively the community has embraced testing and adopted the testing frameworks. Lastly, RQ3 dives deeper into the specifics of how developers write unit tests for open-source DL projects.
To answer the RQs, we conducted an in-depth study of unit testing in open-source DL projects. Based on the IEEE Standard~\cite{iso2017iec} and Keras documentation~\cite{web:Keras}, we categorized unit types in DL models into seven categories: Layer, Loss Function, Optimizer, Activation Function, Metric, Util, and Others. We then analyzed 9,129 open-source DL projects from Github to evaluate the frequency of unit tests. Our findings suggest that projects with unit tests have higher Github popularity metrics, unit tests lead to better project management, and project owners prefer the pull requests with unit tests. In addition, we also provide guidance on thoroughly testing DL project components to identify common flaws.

We summarize the key contributions of this paper as follows:

\begin{itemize}
    \item To the best of our knowledge, this is the first study to analyze the role of unit tests in open-source DL projects, including the benefits unit tests could bring to open-source DL projects and an investigation of DL projects that have adopted unit tests. The study highlights the importance of unit tests in ensuring the reliability and stability of open-source DL projects.
    \item We built up a taxonomy of unit tests in DL projects which contains unit types, tested properties, and assertion sentences. In addition, we integrate our taxonomy with an existing one, linking common DL faults with different unit types in DL projects. The proposed and integrated taxonomies aim to guide the DL model developers in writing test cases.
    \item We systematically collected a dataset for unit testing in open-source DL projects. The dataset may inspire other related research, such as automated test case generation, bug repair, and further empirical studies. 

\end{itemize}

\section{Background}
\label{sec:background}

\begin{figure}
	\centering
	\includegraphics[width=0.7\textwidth]{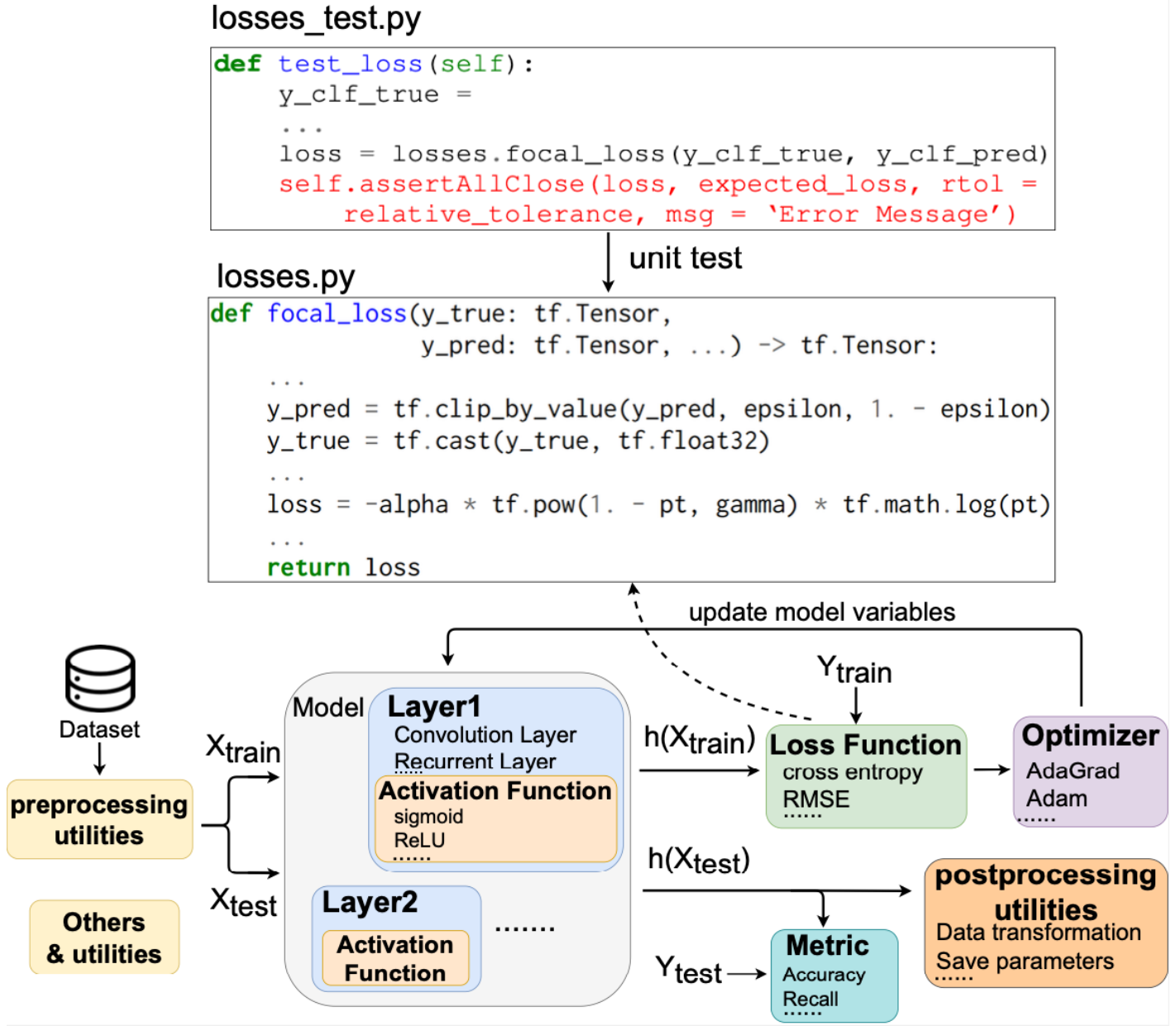}
	\caption{An example code for a loss function with its unit test code and a flowchart showing the units and their interactions in DL projects.}
	\label{fig:DLUnit}
\end{figure}

\textbf{Units in Deep Learning Projects:} 
The IEEE International Standard for Systems and Software Engineering specifies that a unit represents the smallest component that cannot be further subdivided and can undergo testing independently~\cite{iso2017iec}. In the context of object-oriented programming, this typically corresponds to the class level~\cite{binder1999testing}. In the field of DL, some components possess unique roles and require further classification based on their functions. Drawing upon Keras~\cite{web:Keras}, a widely-recognized DL library, we categorize DL model units into seven distinct types.
We illustrate how each part of a unit in a DL project works in Fig.~\ref{fig:DLUnit}. First, the training and testing data are produced by pre-processing utilities based on the input dataset. Second, the Layers in the DL Model calculate and produce the prediction data, where activation functions usually produce the output values of Layers. Third, the Loss function calculates the loss value by comparing the training data labels and predicted value, while the Optimizer updates the model parameters for optimization. The Metric unit calculates the metric value of the model to evaluate its performance. Finally, there may be some post-processing work to be done.

\textbf{Bugs in Deep Learning Projects:} Faults or bugs inside DL projects have been well studied by researchers and may occur in different units of a DL project~\cite{islam2020repairing, zhang2018empirical, humbatova2020taxonomy}. 
Some bugs in DL projects may cause crashes in the model and can be identified easily based on the framework's error message. Nevertheless, Islam et al.~\cite{islam2020repairing} found that 31 out of 50 randomly selected crashed bugs in DL projects either had no messages or the fixes were unrelated to the returned messages. The faults may hide inside the model and won't trigger any errors or crashes, which makes them hard to debug. For example, the user tried to build a classifier to predict the match results~\cite{web:redditQuestion}, but the model always outputs 0\% Loss and 100\% Accuracy with no error messages. In the source code, he defined \textit{tf.nn.softmax\_cross\_entropy\_with\_logits} as the loss function. Due to the nature of softmax, the single output from the logit will always be 1. Hence, after the cross entropy, the output from the loss function will be 0. Such inappropriate usage of loss function bugs can be identified by having a unit test on the output of the loss function value to ensure it will never be 0~\cite{web:whyunittest}. We will discuss mappings between bugs/faults and unit testing in DL projects later in the paper.

\textbf{Unit Tests in Deep Learning Projects:} A unit test is a small and executable piece of code that exercises the functionality of a unit under test~\cite{almasi2017industrial}. It's part of software testing and can detect bugs early.
Based on the IEEE standard~\cite{iso2017iec}, we conclude unit tests in DL projects are testing of individual modules within or related to the DL model by the developer to ensure no errors or bugs in the whole model. For example, in Fig.~\ref{fig:DLUnit}, \textit{losses.py} is the loss function file within an open-source DL project, and \textit{losses\_test.py} is its unit test. Inside the unit test file, mocked parameters like \emph{y\_clf\_true} are initialized and passed into the tested function \emph{focal\_loss()}. Then, the assertion statement checks whether the actual output \emph{loss} is within the expected and acceptable tolerance; if not, the unit test fails, and an explanatory error message is returned.

To support developers and researchers writing unit tests for DL models, popular ML frameworks have introduced built-in unit test functions like tf.test in TensorFlow~\cite{mcclure2017tensorflow}. By default, the unit test framework embedded in Python only provides 33 assertion sentences, whereas the tf.test offers 87 different assertion sentences to fit the unique requirements of DL developers. For example, the \textit{assertAllClose} is used to assert that two structures of lists, NumPy arrays or Tensors, have near values within the tolerance. It has been seen in many of the unit tests in ML/DL projects but is not available in the default test framework. Other assertion sentences like \textit{assertBetween} and \textit{assertArrayEqual} are also commonly seen in DL projects but not used by standard Python projects.

\section{Methodology}
\label{sec:methodology}

This section explains how we collect and analyze the data used in our study. Also, we discuss the motivation and approach for each research question.

\subsection{Data Collection}
\label{sec:datacollection}
\begin{figure}[!ht]
	\centering
	\includegraphics[width=0.65\textwidth]{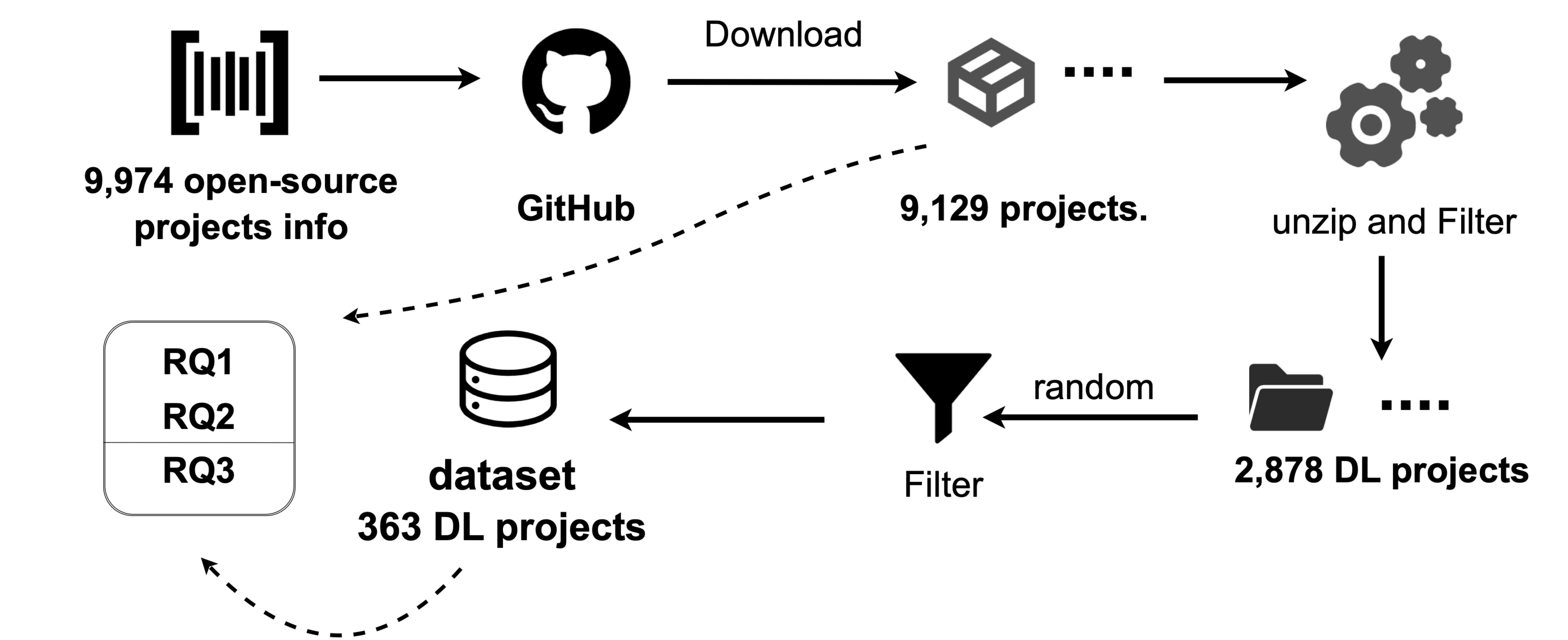}
	\caption{Data collection pipelines for the research questions. }
	\label{fig:data_collection}
\end{figure}

In Fig.~\ref{fig:data_collection}, we present our data collection process. First, we collected open-source projects from Papers With Code~\cite{web:papersWithCodeData} that were implemented using the TensorFlow framework (which is known as the most popular framework~\cite{zhang2019software}) and released after 2017 to ensure that our analysis was up to date. Then, we matched these projects to their respective code repositories and downloaded 9,129 open-source DL projects from GitHub (for RQ1). Third, we filtered 2,878 of them containing unit test scripts based on whether they contained Python files with the string ``test" in the file name or file path ~\cite{trautsch2017there, trautsch2020unit} (for RQ1 and RQ2).
Finally, we randomly sampled 400 DL projects with unit tests. After removing the duplicates and archived projects, we achieved a dataset containing 363 open-source DL projects (minimal criteria size with a confidence level of 95\% and confidence Interval as 0.05~\cite{zar1999biostatistical}) with unit tests (for RQ2 and RQ3). The projects come from various sources, including ready-to-use frameworks, academic research works, and industry projects from large tech companies(e.g., Google, IBM, LinkedIn, and Uber).

\begin{table*}[!ht]
\centering
    \resizebox{1\textwidth}{!}{%
\begin{tabular}{l|l|l}
\hline\hline
                                        \textbf{Name}                                      & \textbf{Category}                                               & \textbf{Definition}                                                                                                                                                                                                                                                                                                                  \\ \hline
\multicolumn{1}{c|}{}                                                         & \textbf{Loss Function}\cite{web:Keras}                                          & The purpose of loss functions is to compute the quantity that a model should seek to minimize during training.                                                                                                                                                                                  \\
\multicolumn{1}{c|}{}                                                         & \cellcolor{gray!25}\textbf{Optimizer}\cite{web:Keras,choi2019empirical}                      & \cellcolor{gray!25}\begin{tabular}[c]{@{}l@{}}Optimization algorithms are typically defined by their update rule, which is controlled by hyperparameters\\ that determine its behavior (e.g. the learning rate).\end{tabular}    \\
\multicolumn{1}{c|}{}                                                         &                                                        & Activation functions are functions used in neural networks to compute the weighted sum of input and biases,                                                                                                                                                                                                               \\
\multicolumn{1}{c|}{}                                                         & \multirow{-2}{*}{\textbf{Activation Function}\cite{web:Keras, nwankpa2018activation}}                  & and  transform them to the output of a neuron in a layer.                                                                                                                                                                              \\
\multicolumn{1}{c|}{}                                                         & \cellcolor{gray!25}\textbf{Layer} \cite{web:Keras}                           & \cellcolor{gray!25}A layer consists of a tensor-in tensor-out computation function and some state, held in variables.                                                                                                      \\
\multicolumn{1}{c|}{}                                                         & \textbf{Metric} \cite{web:Keras}                                                & A metric is a function that is used to judge the performance of the model.                                                                                        \\
\multicolumn{1}{c|}{}                                                         & \cellcolor{gray!25}                               & \cellcolor{gray!25}Utilities used in DL projects, such as model plotting utilities, data loading utilities, serialization utilities,                                                                                                                                                          \\
\multicolumn{1}{c|}{}                                                         & \multirow{-2}{*}{\cellcolor{gray!25}\textbf{Util}\cite{web:Keras} }        & \cellcolor{gray!25}NumPy utilities, backend utilities.                                                                                                                                                                                               \\
\multicolumn{1}{c|}{\multirow{-9}{*}{\textbf{DL units}}}                                & \textbf{Others   }                                              & Other units in DL projects,  such as user-defined label classes, sampling, functions.                                                                                                                                                                                               \\ \hline
                                                                              & \cellcolor{gray!25}                               & \cellcolor{gray!25}The input/output (I/O) test contains tests for value range, shape, type, etc. It can detect whether the input data meets the                                                                                                                                                      \\
                                                                              & \multirow{-2}{*}{\cellcolor{gray!25}\textbf{Input/Output}\cite{islam2019comprehensive,web:unittest}} & \cellcolor{gray!25}model input requirements, and also reflects whether the functions implemented in the unit meet the expectations.                                          \\
                                                                              &                                                        & The error-raising test refers to whether the code can correctly throw an exception, and is often used to check                                                                                                                                                                                                         \\
                                                                              & \multirow{-2}{*}{\textbf{Error-raising}\cite{islam2019comprehensive,web:unittest}}                        & whether the magnitude range of the values generated during the calculation meets the design requirements.                                                                                                     \\
                                                                              & \cellcolor{gray!25}                               & \cellcolor{gray!25}The metric test consists of a test of the metric calculation function and the output metric values after the model                                                                                                                                                         \\
                                                                              & \multirow{-2}{*}{\cellcolor{gray!25}\textbf{Metric}}       & \cellcolor{gray!25}is trained. It is often used to check whether the metric calculation process is correct or the model design is valid.                                                                                                              \\
                                                                              & \textbf{Config\cite{islam2019comprehensive}  }                                                & The config test contains tests for configurations to check whether the unit has been initialized correctly.                                                                                                                 \\
                                                                              & \cellcolor{gray!25}                               & \cellcolor{gray!25}Variable testing includes the range of values, shape, gradient of variables, etc. It can reflect whether the                                                                                                                          \\
                                                                              & \multirow{-2}{*}{\cellcolor{gray!25}\textbf{Variable}}     & \cellcolor{gray!25}model structure is designed as expected and whether the model parameters can be updated.                                                                                                \\
\multirow{-10}{*}{\begin{tabular}[c]{@{}l@{}}\textbf{Unit}\\ \textbf{properties}\\ \textbf{to be tested}\end{tabular}} & \textbf{Others}                                                 & Other tests, for example,  test whether files are generated, whether web requests are successful, etc.                               \\

\hline\hline
\end{tabular}
}
    \centering
	\caption{Types of deep learning units and properties to be tested that are discussed in our study. }
	\label{tab:RQ1Table}
\end{table*}

\subsection{Study Design}
\label{sec:rq_method}
\textbf{RQ1: How can unit tests help open-source DL projects?}
To understand how having unit tests will help open-source DL projects, we performed quantitative analysis over all of the 9,129 GitHub DL projects. With the support of GitHub API~\cite{web:githubAPI}, we managed to crawl the basic metrics of an open-source project on GitHub, which includes issues, pull requests, contributors, stars, and forks. We also calculated the project size by KLOC with the \textit{cloc} tool~\cite{web:cloc}. We used the Mann–Whitney U test~\cite{mann1947test} and Cliff's Delta effect size~\cite{cliff1993dominance}, to calculate the sample differences. To examine the correlation, we applied Pearson’s correlation~\cite{stigler1989francis}. Following the normalizing Github metrics procedure introduced by Jarczyk et al.~\cite{jarczyk2014github}, we applied logarithmic transformation $x' = log_{10}(x + 10)$ to the metrics before calculating the correlation. 

Furthermore, we explored the issues and pull requests of the projects for more details. For each issue in a DL project, we extracted the associated labels. Initially, we organized these based on the 9 standard labels provided by GitHub~\cite{web:githubManageLabels}, resulting in 4 principal categories. Subsequently, the first two authors undertook manual labelling for labels recurring over 100 times, leading to a total of 6 distinct categories: Bug, Dependency, Question, Document \& Enhancement, Status, and Other. For the pull requests in a DL project, we first determine if they contain unit tests by checking the changed file list. Then, we read the status of a pull request and recorded its detailed description content with comments. For the DL projects with unit tests, we compared the PR acceptance rate of the PRs with unit tests and the PRs without unit tests. The acceptance rate is derived by dividing the number of accepted unit-tested PRs by the total number of unit-tested PRs, and vice versa. In addition, we also calculate the odd ratio. The odds ratio is computed by dividing the odds of PRs with unit tests being accepted by the odds of PRs without unit tests being accepted. A value greater than 1 suggests that having unit tests is associated with a higher likelihood of PR acceptance.

In Section~\ref{sec:rq1}, we analyzed the results of DL projects that utilized unit tests and those that did not, using the data we collected. Although the findings from RQ1 may seem trivial, it is important to present empirical evidence for the benefits of incorporating unit testing in DL projects, as there is currently a lack of research on the subject.

\textbf{RQ2: To what extent are DL projects unit tested?}

To investigate the prevailing state of unit tests in the DL domain, we examined 9,129 projects to determine the use of unit tests for DL programs. For a deeper understanding, we specifically utilized the 363 DL projects that employ unit tests to explore the associated files, testing frameworks and the assert statements utilized by developers.

For the presence of unit tests in open-source DL projects, we refer to recommendations on writing unit tests from the official TensorFlow~\cite{web:tensorflowUnittest} and Python~\cite{web:unittest} websites. Moreover, Trautsch et al.~\cite{trautsch2020unit,trautsch2017there} and Yu et al.~\cite{yu2022automated} discussed filtering rules to check unit tests in Java and Python projects. Based on these works, we identify unit tests in DL projects by ascertaining whether ``test'' exists in the filenames or file paths of DL projects.
For example, given a source code file \textit{foo.py}, developers always name its unit test as \textit{foo\_test.py}, or put all the test script files in a separate test folder with the name \textit{foo\_test}. 

For the unit-tested projects, we further investigated the files associated with tests, testing framework frequency, and the assertions being used. Given the complexities of configuring and running unit tests for the vast number of projects, we took a static analysis approach for the 363 unit-tested projects. Specifically, we analyzed the import statements within the unit test files to estimate the number of source code files they're associated with. By dividing this number by the total file count, we derived an approximation of the test coverage in terms of files. 
To learn the frequently used testing frameworks, we used regular expressions to extract the import statements and assertion sentences from the test script files in each project (from RQ1) and sorted them by the number of occurrences. The first two authors then manually identified the test frameworks and checked the framework documentation and the API usage instructions to ensure that the framework provides testing and assertion functionalities.

\textbf{RQ3: Which units and properties are tested in DL projects?}

We analyzed the 363 open-source DL projects with unit tests to understand which units of the DL models are tested frequently and which units need more attention. In addition, we looked into what properties are tested, i.e., what contents are tested in each unit test, in order to gain some insight into unit tests in DL projects. Due to the large number of test files (over 6k) in our dataset, we adopted automated classifiers to categorize the unit testing cases in DL projects.

In our effort to classify unit types within DL projects, we initially consulted the Keras API documentation~\cite{web:Keras}, given Keras's widespread adoption and its representation of high-level DL project abstractions~\cite{humbatova2020taxonomy, moolayil2019learn}. The API categories in Keras, ranging from Models to Utilities, served as a foundational reference. Post some refinements which involved omitting overly high-level APIs and merging closely related ones, we established seven distinct unit types for DL projects. These unit types are outlined in Table~\ref{tab:RQ1Table}. In addition to these, we identified six commonly tested properties within each unit, drawing insights from a notable study by Islam et al. on Deep Learning Bug Characteristics~\cite{islam2019comprehensive} and from recommendations in the official Python documentation~\cite{web:unittest}

To classify unit types and unit properties within DL projects, we initially consulted the Keras API documentation~\cite{web:Keras}, given Keras's widespread adoption and its representation of high-level DL project abstractions~\cite{islam2019comprehensive, moolayil2019learn}. After some refinements which involved omitting overly high-level APIs and merging closely related ones, we established seven distinct unit types for DL projects. These unit types are outlined in Table~\ref{tab:RQ1Table}. In addition to these, we identified six commonly tested properties within each unit, drawing insights from a notable study by Islam et al. on Deep Learning Bug Characteristics~\cite{islam2019comprehensive} and from recommendations in the official Python documentation~\cite{web:unittest}. Based on the definition and the identified types, the first two authors manually looked through the test cases and noticed that the unit type being tested can be reflected by the test class name or test file name. For example, 
the file \textit{test\_loss.py} in Mxnet~\cite{web:mxnetTest} tests the loss functions, and \textit{test\_lstm\_layer.py} in DeText~\cite{web:detextTest} tests the layer within the model. 
Similarly, we extracted the unit properties from the method names and test parameter names.

We built up a value dictionary for each category defined in Table~\ref{tab:RQ1Table}. Given one category, we referred to the method names collected in Paper With Code dataset~\cite{web:papersWithCodeMethod}. For example, we checked the activation function names listed on the website and included them in the activation function category, e.g., relu, tanh. 
We developed two classifiers to categorize DL units and properties based on a given DL test file. These classifiers utilize the Abstract Syntax Tree (AST) to obtain a structured representation of the source code~\cite{baxter1998clone}.  By traversing through each node of the tree with the $NodeVisitor$ provided by the AST library in Python, the classifiers can extract the class, function, and method calls. We override the $nodeVisitor$ function in AST to record the method calls.
For example, to extract properties, our override of the $visit\_{Call}$ function. Within this function, we check for the presence of the string ``assert" in the call name. If found, we extract the name of the tested parameter from the assertion statement using the $args$ attribute. After extraction, the classifiers match these names against our predefined dictionary, which in turn, determines the appropriate categories for each file.

To evaluate the accuracy of our classifiers, we randomly selected 30 files from each category. Then, the first two authors of this paper manually checked if the predicted results were correct individually. Finally, they got together and worked out any differences until they reached an agreement.

We evaluated the unit type classifier and the unit property classifier separately. The unit type classification is a multi-class problem (one object can only be assigned to one non-binary category). Based on our calculation, the average accuracy of our classifier is 0.89, precision is 0.89, recall is 1, and F1 score is 0.94. Note that for each category, we randomly selected the ones that are predicted as true in that category, therefore, the precision value is the same as the accuracy and the recall value is always 1. 

The classification of unit properties to be tested is a multi-label problem (one object can be assigned to multiple labels).
We evaluate the results in both label-based and example-based ways~\cite{zhang2013review}.
Label-based evaluation is similar to how we evaluate the previous classifier for unit types (randomly selected 30 files from each category). The average accuracy is 0.89, precision is 0.89, recall is 1, and F1 score is 0.94. To calculate the example-based metrics, we randomly select a subset of 30 files from all test files and calculate the $Subset Accuracy$ and $Hamming Loss$ metrics. 
The results are as follows: Subset Accuracy is 0.77, Hamming Loss is 0.056, Accuracy$_{exam}$ is 0.86, Precision$_{exam}$ is 0.95, Recall$_{exam}$ is 0.88, and F1$_{exam}$ is 0.91. Details of the evaluation data are released online~\cite{web:unittestinDL}. 

With the two automatic classifiers, we could identify the unit types and what properties were tested in the given test cases. Therefore, we built up a taxonomy of unit tests in DL projects based on the results from the classifiers. First, we categorized the test cases into different unit types by the unit type classifier. Second, we applied the unit property classifier to the test cases under each of the unit types separately. Third, we count the occurrence of the assert sentences under each category with AST. With such data, we construct the taxonomy of unit tests in DL projects, including unit types, tested properties, and frequently used assertions. Note that due to the space limit, we only pick the top 3 tested properties and some unique assertion sentences under each unit type in DL projects. We will discuss the taxonomy in detail in Section~\ref{sec:rq3}.

\section{Results}
\label{sec:results}

\subsection{RQ1: How can unit tests help open-source DL projects?}
\label{sec:rq1}

\begin{figure*}
\centering
\subfigure
[Average Star Count Grouped by KLOC] 
{\includegraphics[width=4.5cm]{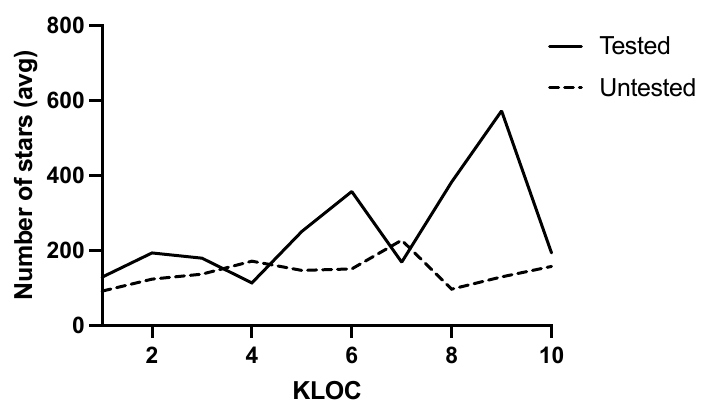}} 
\subfigure[Average Issue Count under All States Grouped by KLOC]{\includegraphics[width=4.5cm]{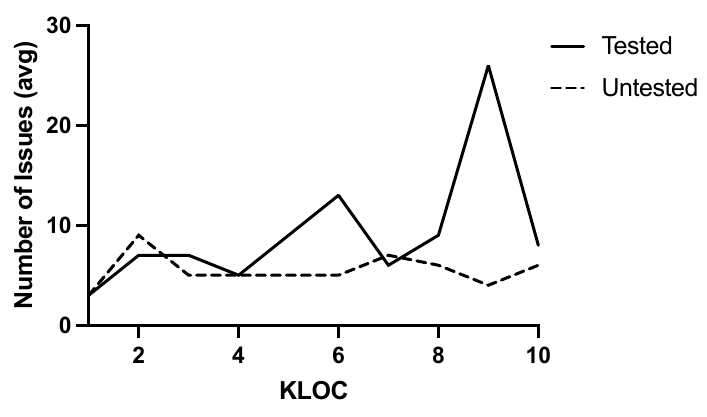}}
\subfigure[Average Fork Count Grouped by KLOC]{\includegraphics[width=4.5cm]{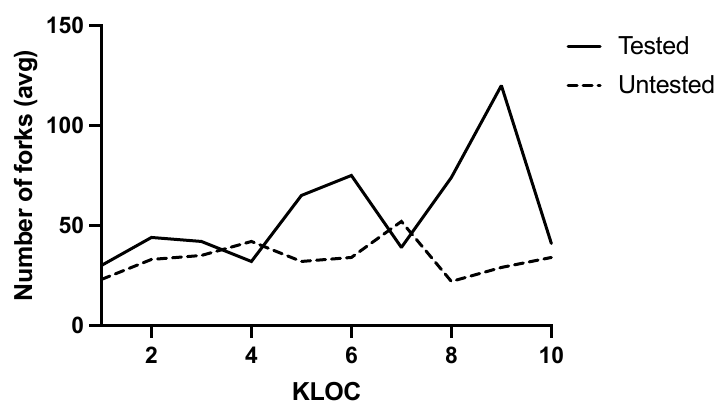}}
\subfigure[Average PR Count Grouped by KLOC]{\includegraphics[width=4.5cm]{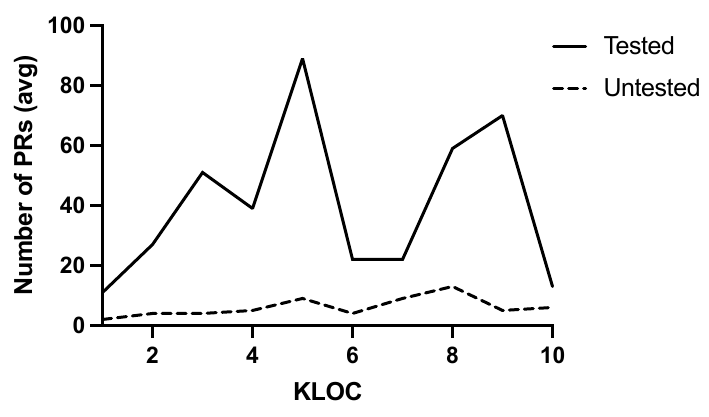}}
\subfigure[Average Individual Contributors Count Grouped by KLOC]{\includegraphics[width=4.5cm]{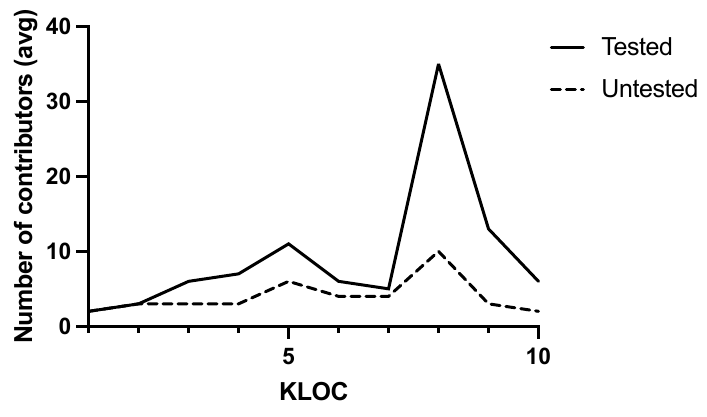}}

 \caption{A comparison of Github Metrics between the unit-tested open source DL projects and the untested projects.} 

\label{fig:rq1} 
\end{figure*}

\textbf{\textit{Finding 1:} Unit tests in open-source DL projects correlate with the number of stars, forks, and contributors.}

In this study, we evaluate the popularity of an open-source project by five different characteristics, which are stars, forks, issues, pull requests, and contributors, inspired by previous works~\cite{borges2016understanding, tsay2014influence}. Table~\ref{tab:correlation} presents the group difference and correlation results. As a result, unit-tested projects have statistically significantly higher characteristics ($p < 0.01$) for all five categories with (small) effect sizes ranging from 0.15 to 0.33. The presence of unit tests demonstrates a positive correlation with all five metrics, though not all correlations are strong. Specifically, for the number of pull requests ($p=0.41$), we observe a robust correlation, suggesting that unit-tested projects are more closely associated with increased pull request activity.

To account for variations arising from project sizes, we normalized the number of those popularity metrics using the KLOC (thousands of lines of code) of each project. Fig.~\ref{fig:rq1}(a)-(e) compares the five characteristics between all DL projects with unit tests and those without unit tests. Specifically, Fig.~\ref{fig:rq1}(d-e) consistently displays unit-tested projects surpassing the untested ones, suggesting greater engagement from developers in contributing or reproducing to the project. While Fig.~\ref{fig:rq1}(a-c) shows some overlaps between the two groups, unit-tested projects generally have higher average values for stars, issues, and forks.

\begin{table}
	\centering

\begin{tabular}{l|ccc}
\hline\hline
Metrics     & Mann-Whitney U $p$ & Effect Size & Pearson $p$ \\ \hline
Star        & \textless 0.01   & 0.18        & 0.10      \\
Issue       & \textless 0.01   & 0.22        & 0.14      \\
Fork        & \textless 0.01   & 0.15        & 0.10      \\
PR          & \textless 0.01   & 0.33        & 0.41      \\
Contributor & \textless 0.01   & 0.19        & 0.19      \\ \hline\hline
\end{tabular}

	\caption{The sample differences and correlation of popularity metrics with the presence of unit tests}
	\label{tab:correlation}
\end{table}

\begin{figure}
    \centering
	\includegraphics[width=0.45\textwidth]{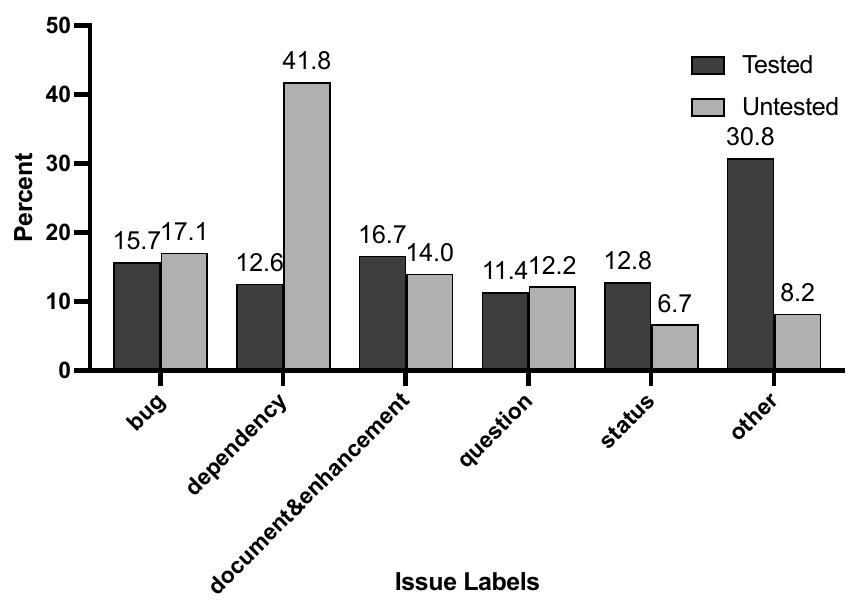}
	\caption{A comparison of Github issue label categories of unit tested and untested DL projects (in percentage).}
	\label{fig:issues}
\end{figure} 

\textbf{\textit{Finding 2:} Pull requests with unit tests have a higher acceptance rate.}

We also crawled the pull requests (PR) details from DL projects that contain unit tests. In total, we collected 76,105 PRs from DL projects with unit tests, where 25\% of the PRs include updates in the unit test files. Similar to the findings by Tsay et al.~\cite{tsay2014influence} in conventional open-source projects, we also find that the inclusion of unit tests influences the PR acceptance rate in DL projects. Specifically, 82.8\% of the PRs with unit tests are merged into the source code, whereas only 74.4\% of those without unit tests. The odds ratio is 1.113, which is higher than the result (odd ratio = 1.059) from the previous study that was conducted across more general conventional projects~\cite{tsay2014influence}. This suggests that while unit tests are associated with a higher likelihood of PR acceptance in general open-source projects, their significance is even more pronounced in open-source DL projects. 
Aside from the acceptance rate, PRs with unit tests also have a shorter waiting time before the merge. We find that, on average, the PRs with unit tests (6.25 days) is 4 hours faster to be accepted than those without unit tests (6.43 days). It could be due to the PR with unit tests that can help the developer automatically identify the potential bugs in the submitted code. Sometimes, the project owner may ask the contributor to add a unit test for the PR or ask them to double-check if the current unit tests are passed before the merge.

\textbf{\textit{Finding 3:} Issues from unit-tested projects are resolved quicker and have more project management labels.}

In addition to finding 1 and 2, we investigated issue management by comparing the closing rate and the time taken to close issues and categorizing different issue types between unit-tested and untested DL projects. Our dataset comprised 10,858 issues from projects without unit tests and 105,430 issues from unit-tested projects. We find that issues from unit-tested projects exhibit a higher closure rate (83.9\%) in contrast to projects without unit tests (61.8\%). Similar to the approach used in Finding 2, on average, issues in unit-tested projects are closed in 42.7 days, which is 4.8 hours faster compared to those in projects without unit tests, where the average closure time is 42.9 days. This shorter time frame for issue resolution and the high issue close rate can indicate that unit-tested projects are able to identify and address issues more swiftly.  This agility contributes to enhancing the project's overall dependability and makes it more reliable~\cite{aversano2015analysing}.

Fig.~\ref{fig:issues} presents the issue label type distribution in a grouped bar chart. Dependency has the most labels in the untested DL project issues (41.8\%). The dependency labels are mainly automatically generated when one of the libraries used by the project is updated.
On the contrary, the issue labels from DL projects with unit tests are typically generated and managed by developers and are more related to status (e.g., approve, duplicated, done) and document \& enhancement (e.g., feedback, upgrade, suggest). Most other issue labels (for DL projects with unit tests) are related to the contributor license agreement (CLA), which benefits the project management and encourages developers to contribute. Besides, we notice more abbreviated labels used in the tested DL projects, e.g., lgtm, wip, tbd. 
Overall, we find that DL projects with unit tests are under better management, one of the advantages that unit testing brings to a software project.

\begin{center}
 \begin{tcolorbox}[colback=black!5!white,colframe=black!75!black,bottom=-0.05pt,top=-0.05pt]
Summary: Our analysis revealed unit-tested DL projects have a positive correlation with open-source project metrics, such as the number of stars, forks, pull requests, and a higher acceptance rate of pull requests. Moreover, unit-tested projects demonstrated faster and more efficient issue resolution. Additionally, they were also found to have more project management labels. These findings suggest that unit tests play an essential role in ensuring the reliability and stability of open-source DL projects and highlight the need for rigorous testing to maintain their quality and prevent issues from propagating downstream.
\end{tcolorbox}
\end{center}

\subsection{RQ2: To what extent are DL projects unit tested?}
\label{sec:rq2}
\textbf{\textit{Finding 4:} 68\% of the DL projects do not have unit tests and the unit test file coverage is low.}

\begin{figure}
    \centering
	\includegraphics[width=0.45\textwidth]{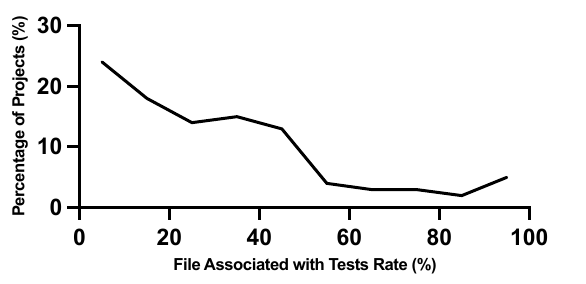}
	\caption{The distribution of open-source DL projects based on their file associated with tests rate.}
	\label{fig:fileCoverage}
\end{figure}

We present the results from RQ2 in this section to provide a general picture of how unit tests are deployed in the open-source DL projects.
31.52\% of the open-source DL projects in our dataset contain unit tests, whereas the rest of the projects (68.48\%) do not have unit tests.
It can be due to that some of the projects are from researchers who prefer to focus on the results compared to the code testing~\cite{yao2017unit}. The lack of testing in such DL projects may make it harder to ensure the code is working as intended and producing accurate results, present barriers for other developers who want to replicate and improve the original model, and also hinder the growth of open-source projects.
Additionally, we also notice some of the untested projects are simply the Tensorflow version of a model that was originally developed by other frameworks. Third, there are also developers who choose to use .ipynb type of file for their model, which needs extra effort to test the files~\cite{web:testJupyter}. Fig.~\ref{fig:fileCoverage} demonstrates the distribution of open-source DL projects based on their file associated with the tests rate. 56\% of the projects have less than 30\% of their files associated with test files. And only 5\% of the projects have a coverage over 90\%. This suggests that while many DL projects employed unit tests, a substantial number of files have not been covered by the tests.

\textbf{\textit{Finding 5:} Numpy and Tensorflow are the most commonly used testing frameworks.}

In Table~\ref{tab:testFramework}, we present the usage of testing frameworks in DL projects.
The NumPy~\cite{web:numpyTesting} and TensorFlow frameworks are used the most for writing unit tests, accounting for 77.6\% and 64.7\%, respectively. Some DL projects use both frameworks. In some simple test scenarios, the basic assertion methods provided by NumPy are more straightforward and easier to use, or developers need other methods than NumPy provides to initiate parameter values, e.g., \textit{np.int32}, \textit{np.random}, and \textit{np.zeros}. If more complex tests are needed, or if the test environment relies on the TensorFlow run-time environment, more developers will choose the tools provided by TensorFlow for testing~\cite{web:tfTest}. In addition, some developers write unit test scripts using frameworks such as unittest~\cite{web:unittest}(the one provided by default in Python) and pytest~\cite{web:pytestTesting}, which provide similar testing methods.

On the other hand, 9.3\% of the DL projects use different frameworks when writing unit tests. Specifically, some developers used less frequent frameworks, custom testing tools, or print functions to write unit tests. For example, the \textit{driver-check.py} in msc-graphstudy ~\cite{web:testDriverCheck} combines if-else and try-except statements using print and raise to output the results of the process and the final prompt message to determine whether the test passes. In some exceptional cases, common testing frameworks may not be utilized for various reasons, such as limited testing knowledge of the developers or unsuitability for the specific task (e.g., when testing whether a library is imported correctly or configuring the model environment).

\begin{table}
	\centering

	    \resizebox{1\textwidth}{!}{
\begin{tabular}{l|llllllll}
\hline\hline
\textbf{Framework}          & Numpy          & Tensorflow     & unittest     & PyTest      & absl       & GoogleTest & nose       & Others       \\ \hline
\textbf{Num. of Projects} & 2,232 (77.6\%) & 1,862 (64.7\%) & 440 (15.3\%) & 284 (9.9\%) & 98 (3.4\%) & 23 (0.8\%) & 19 (0.7\%) & 269 (9.3\%)\\\hline\hline
\end{tabular}
}
	\caption{ Python unit test frameworks used in sampled DL projects (note that one project may include more than one framework).}
	\label{tab:testFramework}
\end{table}

\begin{figure*}
    \centering
	\includegraphics[width=1\textwidth]{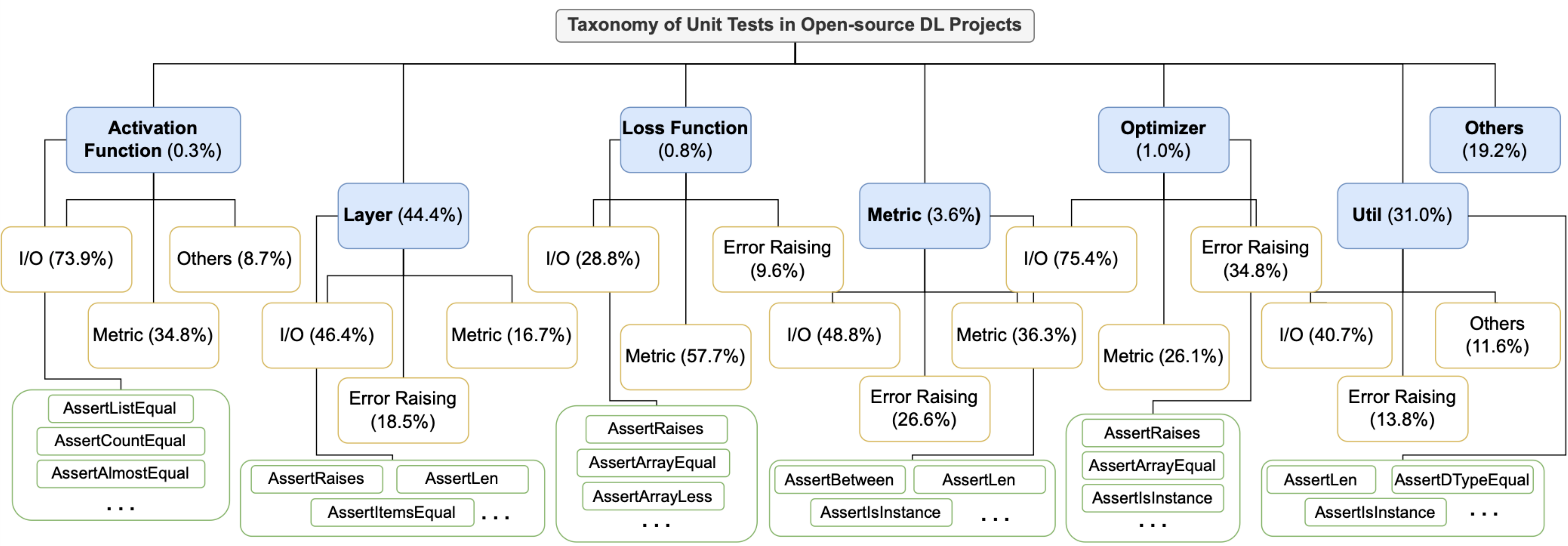}
	\caption{A taxonomy of unit tests in open-source DL projects includes test types, properties to be tested, and frequently used assertions. }
	\label{fig:rq2_result}
\end{figure*}

\textbf{\textit{Finding 6:} 21\% of the assertion sentences are not identified within the dataset.}

We also count the usage of assertion sentences from TensorFlow. As a test framework, \textit{tf.Test}~\cite{web:tfTest} provides 87 assert sentences. 79\% of the assert sentences are used in our dataset. The five most frequently used sentences are \textit{assertEqual}, \textit{assertTrue}, \textit{assertAllEqual}, \textit{assertAllClose}, and \textit{AssertRaises}. Note that \textit{assertAllEqual} and \textit{assertAllClose} are not provided by the unit test framework embedded in Python. They were designed to assert the NumPy ndarraies or Tensors commonly used in DL projects. Besides, some assertions are not found in the dataset. Some of them are the negative form of other assertion sentences used in the dataset, e.g., \textit{assertNotStartWith}, \textit{assertNotRegex}, \textit{assertNotAlmostEquals}. Instead of calling those negative forms, developers use positive forms, such as \textit{assertRegex} and \textit{assertAlmostEquals}. 
Other assertions not found in the dataset may need more attention to be used. 
For example, the \textit{assertWarns} and \textit{assertWarnsRegex} are not used probably due to developers care more about errors than warnings. 
Assertions (e.g., \textit{assertContainsExactSubsequence}, \textit{assertContainsInOrder}, and \textit{assertContainsSubsequence}) that check the subset of a string or list may be created for uncommon DL tasks or introduced directly from assertions in Java that have similar functionalities.

\begin{center}
 \begin{tcolorbox}[colback=black!5!white,colframe=black!75!black,bottom=-0.05pt,top=-0.05pt]
Summary: 68\% of the open-source DL projects in our dataset do not have unit tests. And 56\% of the tested projects have less than 30\% of the files associated with tests. The most common frameworks in the tested DL projects are NumPy and TensorFlow. Special scenarios may have custom testing tools to write unit tests. Some of the less common assertion sentences could require some attention.
\end{tcolorbox}
\end{center}

\subsection{RQ3: Which units and properties are tested in DL projects?}
\label{sec:rq3}

After having a general picture of unit testing in open-source DL projects, we looked into the content of the unit test scripts. We developed classifiers to tell which units are tested and what properties of units are tested in DL projects.

\textbf{\textit{Finding 7:} A taxonomy that links the unit types and tested properties in DL projects.}

A detailed taxonomy of unit types and tested properties is shown in Fig.~\ref{fig:rq2_result} which has been derived by following the process mentioned in Section~\ref{sec:rq_method}.  It presents seven types of units in open-source DL projects. For each unit, the number next to its name represents the occurrence of it among all test files in percentage. It also links with the frequently tested properties and the assertion sentences often used to test the specific unit. Table~\ref{tab:whatTested} presents the overall distribution of properties to be tested in all units. We discuss each of the units in Fig.~\ref{fig:rq2_result}.

\begin{table}
	\centering
            \resizebox{1\textwidth}{!}{
\begin{tabular}{l|llllll}
\hline\hline
\textbf{Framework}          & Input/Output   & Error Raising  & Metric         & Config      & Variable  & Others       \\ \hline
\textbf{Num. of Test Files} & 3,467 (50.3\%) & 1,223 (17.8\%) & 1,005 (14.6\%) & 538 (7.8\%) & 345 (5\%) & 907 (13.2\%)
\\ \hline\hline
\end{tabular}
}
	\caption{Properties of units tested in DL projects in terms of test files from the dataset. }
	\label{tab:whatTested}
\end{table}

\begin{figure}[!ht]
\begin{minted}{python}
import tensorflow as tf
def test_conv2d(self):
    ...
    output = cl.conv2d(inputs, 'test_conv2d', 
       filter_size, in_channels, out_channels, strides)
    output_shape = [2, 5, 5, 4]
    self.assertAllEqual(tf.shape(output), output_shape)
\end{minted}
\caption{Example code to test the shape of the output value in one layer.}
\label{fig:testLayerOutput}
\end{figure}

\begin{figure}[!ht]
\begin{minted}{python}
def test_net_construct(self, net):
    with self.assertRaisesRegex(ValueError,
            "output_channels must not be empty"):
        net(output_channels=[],
             kernel_shapes=self.kernel_shapes,
             strides=self.strides,
             paddings=self.paddings)
\end{minted}
\caption{Example code to test the layer variable with \textit{assertRaisesRegex}.}
\label{fig:testAssertRaises}
\end{figure}

\textbf{Layer. }
The Layer unit has the most unit tests among all other unit types, as it is the core unit of a DL model. A layer in the DL model is also known as a block or a single net. Layers of neurons form a DL network. Fig.~\ref{fig:testLayerOutput} shows an example of a test case for a 2D convolution layer. It gets the \emph{output} (testing I/O) of the tested layer and compares its shape with an expected output shape. The tests vary from testing the single-layer outputs to testing the results of the entire network within an expected range. During the study, we noticed that some developers also tend to misconfigure the layer on purpose and catch the raised errors with \textit{assertRaises} or \textit{assertRaisesRegex} (testing Error Raising), which prevents developers from using \textit{try/catch}, as shown in Fig.~\ref{fig:testAssertRaises}. 
Therefore, fewer \textit{Exception Handling}, known as one of the bad practices in test codes (test smells), occur in Python compared to Java~\cite{wang2021pynose}. To test the Config or Variable of a layer, developers can call \textit{get\_config} or \textit{get\_variables} and assert the value, respectively. 

\textbf{Util. }
Testing the Util unit includes data processing, test initiation, database utilities, and other utility functions. Because the Util unit has the highest number of functions in a DL project, there naturally exists a higher number of unit tests for it. Since the utility functions have straightforward functionalities, writing unit tests for them is easier than other units in DL projects. For example, when testing matrix reshapes functions such as matrix transposition, developers can quickly create some input matrix and the expected outputs. The \textit{AssertDTypeEqual} can be used to check the ndarray data type.

\textbf{Metric. }
Testing the Metric of a DL project aims to ensure the validation works. We noticed that \textit{AssertBetween} or \textit{AssertNear} is used to measure the metric result within expectation. Aide from that, \textit{AssertIsInstance} is also used to ensure the returned metric is in the correct format for reading and computing, as seen in Fig.~\ref{fig:testMetric}.

\textbf{Optimizer. }
In testing the optimizer, developers tend to train the DL model to compute the actual gradients of the loss value. Besides, we also saw some unit tests testing the optimizer's configuration to ensure the function was set up correctly, as seen in Fig.~\ref{fig:testOpt}.

\textbf{Loss Function \& Activation Function. }
Loss Function and Activation Function units are not tested much, likely due to the developers only testing such functions when they need to create a new approach. In most cases, developers call loss/activation API from well-established DL libraries for which they do not need to write test cases. An example of testing the loss function output value is shown in Fig.~\ref{fig:DLUnit}.

\textbf{Others. }
The Others unit includes testing of self-defined functions, sampling functions, math calculations, whether an HTTP request is successful, etc. The unit tests for this unit are similar to unit tests in conventional software projects.

\begin{figure}[!ht]
\begin{minted}{python}
def test_classification_metrics(self):
    ...
    expected_return = {'precision': np.array[], 
                                  'recall': ...}
    metrics = classification_metrics(ground_truth, 
                                        retrieved)
    self.assertIsInstance(metrics, dict)
    for k, v in metrics.items():
        self.assertIsInstance(v, np.ndarray)
        ...
\end{minted}
\caption{Example code to test the classification metrics.}
\label{fig:testMetric}
\end{figure}

\begin{figure}[!ht]
\begin{minted}{python}
def test_adam_tf(self):
    """Test creating an Adam optimizer."""
    opt = optimizers.Adam(learning_rate=0.01)
    global_step = tf.Variable(0)
    tfopt = opt._create_tf_optimizer(global_step)
    self.assertIsInstance(tfpot, tf.keras.optimizers
                                               .Adam)
\end{minted}
\caption{Example code to test the optimizer.}
\label{fig:testOpt}
\end{figure}

\begin{center}
 \begin{tcolorbox}[colback=black!5!white,colframe=black!75!black,bottom=-0.05pt,top=-0.05pt]
Summary: More than 70\% of the unit tests test the Layer and Util unit in DL projects. Activation Function, Optimizer, and Loss Function are the least tested. More than half of the test cases are testing the I/O. Error Raising is also commonly seen in test cases. Config and Variables of the units are not the focus when testing, which may need more attention in the future. We summarized a taxonomy (Fig.~\ref{fig:rq2_result}) that aims to help developers to write test cases for different units in an open-source DL project.
\end{tcolorbox}
\end{center}

\section{Discussion and Implications}
\label{sec:discussion}

Based on the preceding findings, our study reveals the emergence of the need for unit testing in open-source DL projects; despite the advantages that unit testing could provide for projects, not enough have unit tests. We next discuss some actionable implications for practitioners and researchers.

\begin{figure*}
    \centering
	\includegraphics[width=0.95\textwidth]{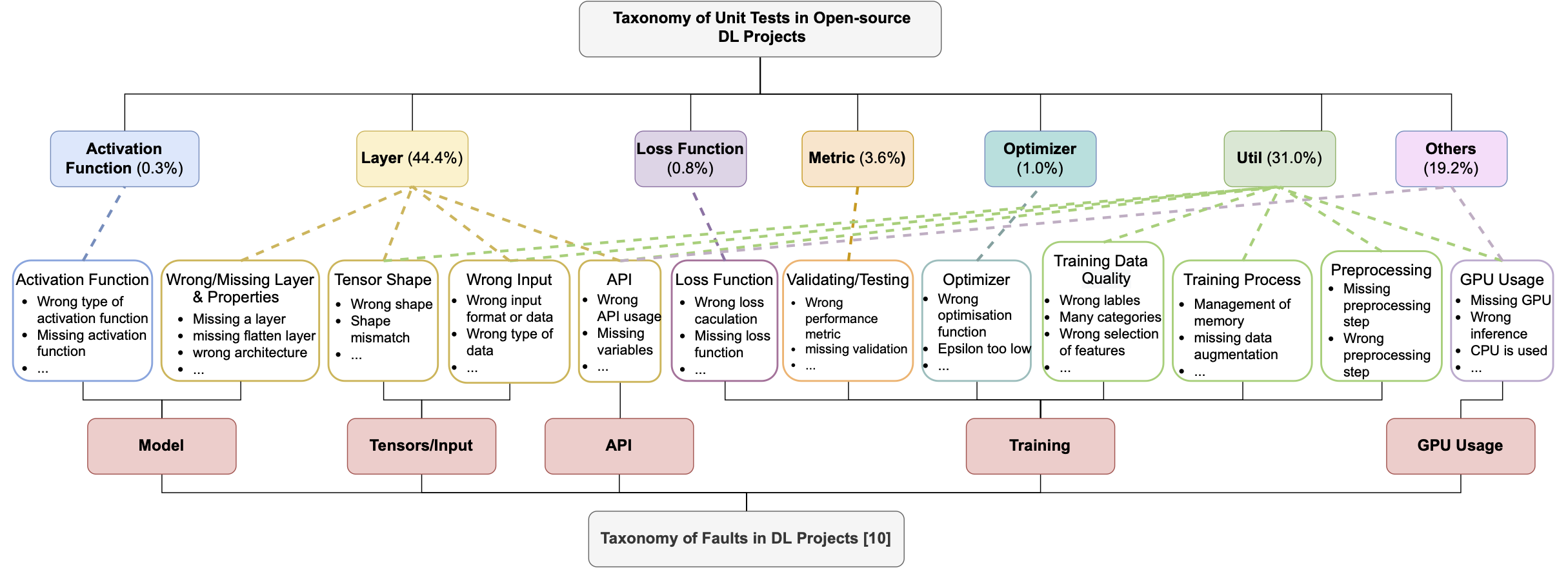}
	\caption{Integration with existing literature~\cite{humbatova2020taxonomy}, the mapping of our taxonomy (unit tests in open-source DL projects) with the taxonomy of faults in DL projects.}
	\label{fig:dis_tax}
\end{figure*} 

\subsection{Implication for practitioners}
Based on the findings in the research questions, we summarise the implications for DL practitioners.

\textbf{Detecting and mitigating Deep Learning faults becomes simpler with a developer's guide in writing DL unit tests.}
In RQ3, we build a taxonomy for unit tests in open-source DL projects to assist developers in writing test cases. Since unit testing is known for identifying faults/bugs inside the code~\cite{shamshiri2015automatically}, we discuss the mapping between DL faults and unit tests in this section.
Previously, Humbatova et al.~\cite{humbatova2020taxonomy} have identified faults in DL systems and built a taxonomy as one of the research outcomes. 
Integrated with their outcomes, we managed to build a taxonomy that maps the deep learning faults~\cite{humbatova2020taxonomy} and unit tests (RQ3), which may help the practitioners prevent the corresponding faults.
The first two authors conducted open coding to build the taxonomy to achieve this. First, they went through each subsection of the faults in open-source DL Projects and tried to find corresponding samples in the code base. Second, each of them conducted the mapping individually. Some of the mappings are intuitive. For example, map the faults in the optimizer with the unit tests in the optimizer. Finally, they met up and solved the disagreements until they reached a consensus. we calculated the Krippendorff's alpha score~\cite{hayes2007answering} and the value ($\alpha=0.87$) indicates a good agreement. 

In Fig.~\ref{fig:dis_tax}, we present the taxonomy of unit types of DL projects mapped with the faults that unit tests could identify. We put the detailed faults into bullet points under the second level of the taxonomy. We hide some faults with dots due to the space limit. A full version of the taxonomy is presented online~\cite{web:unittestinDL}. Note that some faults can be tested by multi-units of DL projects, e.g., Wrong input, API, and GPU Usage.

With such connections, we aim to give developers a guide into how each type of unit test can navigate or prevent certain types of bugs from happening. Given a particular bug, developers shall locate the bugged unit from this taxonomy so that they can refer to Fig.~\ref{fig:rq2_result} to construct a test case to avoid future issues within the unit.
Specifically, to test a missing layer fault, developers can first identify whether the bug is related to the layer unit. Then, they could check the commonly used assertions in the Layer unit in Fig.~\ref{fig:rq2_result}, and assert the length of layers is under expected (e.g., $assertLen(model.layers, num\_layer)$). During the training phase, which has the majority number of faults, we present a sample (Fig.~\ref{fig:DLUnit}) on how to test a self-defined loss function to ensure the output is under expectation by using the $AssertAllClose$ to avoid the wrong loss calculation error. The Validating/Testing faults can be detected by writing unit tests for the Metric, e.g., checking if the metric value is within expectation with $AssertBetween$. The Utility tests can address faults in the Training Data Quality. Before reading the data as input, developers could write unit tests to check if the data is labelled well, which includes asserting the value is equal to the expected ones or if the labelled data size is correct. For the faults in GPU Usage, Tensorflow has provided several APIs to check the number of available GPUs and their names. We found that some unit tests will check the number of GPUs ($tf.config.experimental.list\_physical\_devices('GPU')$) before the training to check whether it is enough for the training or simply check if the GPU is available.

Despite the discussed faults that can be identified by unit testing, we found some faults that have not been well-tested in DL projects. For example, most unit tests in activation function only focus on testing if the function is working as expected, e.g., the correct output value. Few of the unit tests test whether the selected activation function is correct. It may be trivial for the seniors, but for novice developers who tried to reproduce other models, an assert on the activation function name or the outputs from different functions can save them time to debug since the wrong type of activation function is the main fault in this field~\cite{humbatova2020taxonomy}.
Besides, in the Training process, we did not see many tests on the suboptimal number of epochs and batch size. In the faults in the Model, a common fault, namely suboptimal network structure (using too few or too many layers), is also not seen in the unit test cases. This could be because the developers are familiar with their model and will not make such mistakes. However, for the ones who want to reuse the model or conduct some research on it, a suboptimal fault could cause failures. Therefore, developers should consider potential contributors when they write unit test cases for their DL projects.

\textbf{Write unit tests when submitting a pull request.} 
In RQ1, we observe that when submitting a PR, the test inclusion leads to a higher acceptance rate and a shorter waiting period. Therefore, when developers tend to submit a PR, we suggest they comment that all existing tests have been passed and new tests have been added for new features. Some examples from the open-source projects are like ``\textit{List of changes: ... Add unit tests for the metrics module.}" and ``\textit{So, we finished with last updates of code to pass all tests. The last thing is failing and also fixed.}". Alternatively, the project owner may request you to do so, e.g., `\textit{This is an excellent PR. Can you add tests for it?}`". In addition, we also encourage practitioners to submit PRs that are purely related to unit tests (all changed files are test files), especially to projects without unit tests. The requests may be related to fixing test smells in the current unit tests or adding new tests for the existing methods.

\subsection{Implication for researchers}

\textbf{Qualitative studies with practitioners}
While our current study is grounded in a static quantitative analysis of mined repositories, there exists an avenue for further research that can provide a qualitative study in this field. To augment our findings, researchers could conduct on qualitative studies involving direct engagement with DL practitioners. Conducting interviews, surveys, or focus groups with developers could provide deeper insights into their unit testing practices and more challenges. By validating and refining the taxonomies derived from our quantitative study through these interactions, researchers may be able to unearth more aspects of unit testing practices in the context of open-source DL projects and add more to the taxonomies. This approach would enhance understanding of the interaction between unit testing and the open-source DL project development.

\textbf{Support unit tests completion for DL projects}
Though previous works~\cite{yan2019artdl, guo2018dlfuzz,tian2018deeptest,riccio2020model, pei2017deepxplore, ma2018deepgauge} have been conducted on better testing DL projects, a gap remains in the realm of unit testing for such projects. The observed low adoption and file coverage rates for unit tests, as highlighted in RQ2, underscore a need for researchers to delve deeper. It's important to understand the underlying reasons why developers are not writing unit tests and to unearth the challenges developers face. Pinpointing these challenges can pave the way for solutions that facilitate more effective unit test writing, for instance, unit test completion for DL projects. Considering the complexity of test case generation in Python~\cite{lukasczyk2020automated}, an initial step could be assisting developers in completing the assert statements for DL projects. Our findings from RQ3, detailing frequently tested properties across various unit types and preferred assertions (some potentially DL-specific), offer a foundation. Leveraging this, future research can employ tools like Large Language Models to further assist developers in refining and completing their test cases.

\textbf{Enhanced unit test quality in open source DL projects}
With our deep understanding of DL unit types and their associated bugs, there emerges an opportunity to innovate methods that enhance the quality of unit tests in DL projects, ensuring they are comprehensive and address potential vulnerabilities. Similar to Jia et al. ~\cite{jia2022injected} did to the DL libraries, researchers could inject a known type of bugs in DL projects and check if the unit test can detect the certain bug and how the unit test can be improved. For example, given a unit test on the layer of a model, one new approach could be to intentionally inject a bug (e.g., omit the flatten layer in the source code). The test should ideally flag this omission. If the test doesn't catch this bug, there's an opportunity for researchers to refine the test's sensitivity. Beyond just detection, the challenge then extends to providing meaningful error messages or drawing from the taxonomy to suggest specific updates for bug-specific test cases. Such refined testing methods could boost the reliability and robustness of tests in open-source DL projects.

\textbf{Research DL projects also need unit tests}
As mentioned in our study,  only 31.5\% of the projects we collected have unit tests (RQ2). The reasons why open-source DL projects do not have unit tests may be because, for AI researchers, the model may call some third-party APIs and do not think unit tests are needed.
While some DL models could only be composed of a few API calls from DL frameworks, there might still be undetected faults in those calls (seen in Fig.~\ref{fig:dis_tax} and ~\cite{humbatova2020taxonomy}). Having unit tests ahead can reduce the debugging time.
Aside from that, we find that unit-tested open-source DL projects positively correlate with open-source project metrics (RQ1), so researchers should consider adding unit tests for their projects to increase the trustworthiness and reliability of the code and make it more useful for other researchers and practitioners who may also want to reproduce your work in other areas. 

\section{Threats to Validity}
\label{sec:threat}
\textbf{Internal validity} A potential threat to the validity of this study is the detection of unit tests in a project. Although there are official documents~\cite{web:tensorflowUnittest, web:unittest} that guide how to write unit tests, some developers may not follow these guidelines when writing unit tests. This could lead to the potential misclassification of projects. To mitigate this concern, we not only relied on official documents but also reviewed other research~\cite{trautsch2020unit} on unit testing to determine how to identify unit tests in a project. Additionally, we tried to be as thorough as possible in detecting unit tests in a project, but there may be some cases where unit tests are not detected.

Another internal threat is how we defined unit types in DL projects and the properties. To mitigate the potential bias, we refer to the IEEE definition of unit test, Keras documentation, and some related research in DL projects~\cite{iso2017iec, web:Keras, islam2019comprehensive, web:unittest}. 
To mitigate the threat during the evaluation of our scripts (the classifiers for which units and properties are tested), we have two authors independently label the output from the scripts, and the accuracy is fair. During the procedure of building the taxonomy (Fig.~\ref{fig:dis_tax}), manual efforts are required for the integration. To mitigate this, we have two authors (with DL experience) to conduct the mapping and resolve the conflicts for the final results. In addition, the metric to measure the inner agreement is high, demonstrating the result's reliability.

\textbf{External validity} The threats to external validity in our study are related to the data selection and generalization. 
This study collected data from Papers with Code, which contains various open-source DL projects from different publications. Still, it may not be representative of all open-source DL projects that exist. Our investigation of the downloaded open-source DL projects revealed that the dataset included projects developed by academicians and notable projects from tech companies such as the BERT model~\cite{devlin2018bert} or DeText~\cite{guo-liu20}. Recent research~\cite{yang2023users}, using data from Papers with Code, also confirms both sectors' presence, highlighting close collaboration as the industry uses academic frameworks and academics employ industry DL libraries. However, we acknowledge that it may not include projects that do not have publications. This limitation should be taken into consideration when interpreting the results of the study. Furthermore, we recognize that researchers may not always write unit tests for research projects, which could influence the \textit{Finding 4} of this study. Nevertheless, unit testing is still considered a crucial step in the software development process, and this study aims to provide valuable insights into best practices for DL model testing.

\textbf{Construct validity} In RQ1 of this study, we measured the popularity of the open-source projects using the number of stars, forks, and contributors as proxies, which aligns with the methods used in previous studies that have used similar metrics to assess popularity~\cite{borges2016understanding}. Furthermore, previous studies have also confirmed that the number of contributors is a key factor in the success of open-source projects~\cite{mockus2002two, borges2016understanding}. However, it's important to note that using metrics such as stars, forks, and contributors to measure popularity may not be entirely accurate, as developers may star a project for other reasons, such as using it as a bookmark or simply wanting to follow the update. Additionally, it's difficult to determine a causal relationship between unit testing and popularity, as other factors may contribute to a project's success.

\section{Related Work}
\label{sec:relatedwork}

\textbf{Studies on Unit Testing} Recent works in unit testing doubt whether existing standard definitions of unit testing are still valid in modern software development environments~\cite{trautsch2017there,trautsch2020unit,trautsch2019analysis}. Trautsch et al.~\cite{trautsch2020unit} classified 38,782 test cases as unit and integration tests and used mutation testing to evaluate their defect detection ability. They suggested reconsidering the definitions of unit and integration testing. Trautsch et al.~\cite{trautsch2019analysis} studied 27 Java and Python projects with more than 49,000 test cases. They confirmed that unit tests were better at pinpointing the source of the defect. 
Our study defined the unit types in DL projects based on the IEEE definition and our observations of the Keras library to ensure the result is generalized and objective.

\textbf{Testing in DL Projects} In previous studies, researchers studied the challenges~\cite{sculley2014machine,sculley2015hidden,arpteg2018software} in ML-based software systems. They pointed out from a high-level perspective which parts of machine learning-based software systems need to be tested in practice~\cite{breck2017ml}, such as functions and data, model development, infrastructure, and monitoring tests. Later, software engineering researchers conducted more detailed research on data testing and model testing of DL applications. Breck et al.~\cite{breck2019data} proposed a data verification system to specifically detect anomalies in the data sent to the machine learning pipeline. Regarding model testing, a related topic is a study of generating input data to increase the coverage of neural network models. Researchers also have proposed some new neural network coverage metrics~\cite{pei2017deepxplore,sun2018testing,sekhon2019towards,ma2019deepct}, black-box testing ~\cite{yan2019artdl} and white-box testing methods~\cite{guo2018dlfuzz,tian2018deeptest,riccio2020model}. Specifically, the DeepXplore~\cite{pei2017deepxplore} first proposed neuron coverage criteria to drive test generation.  Ma et al.~\cite{ma2018deepgauge} proposed DeepGauge, a set of multi-granularity testing standards for DL systems designed to portray test platforms in many ways. And Yan et al. ~\cite{yan2019artdl} designed the ARTDL algorithm based on the black box testing method RT (random testing) to generate test cases more likely to cause failure. Guo et al.~\cite{guo2018dlfuzz} proposed DLFuzz, which maximizes the neuron coverage and the prediction difference between the original input and the mutated input by keeping a small input mutation. Most of the research mentioned above has been focused on exploring new approaches to increase the coverage of test cases, tested neurons, and platforms. Our study investigates the unit test in current DL systems, the first work in the area to our best knowledge.

This study is one of the few that examines unit testing in DL projects. Zhang et al.~\cite{zhang2020machine} have highlighted the challenges in dividing DL systems into testable units. They cited two older works that may be useful for writing unit tests for ML programs. However, these works are not suitable for recent DL models. Riccio et al.~\cite{riccio2020testing}, and Song et al.~\cite{song2022exploring}  also studied the current testing practices and emphasized the need for more research on unit testing in learning systems to identify bugs.
Our research aims to contribute to the ongoing efforts to improve unit testing practices in DL projects and give guidance on writing unit tests by studying existing test cases in DL projects.

\section{Conclusion}
\label{sec:conclusion}

Deep learning models have become widely used and essential for various applications. Thus, ensuring open-source DL project methods perform correctly via unit testing is critical to their reliability, efficiency, and robustness.
In this research, we examined unit testing in open-source DL projects. Our findings indicate unit tested DL projects have higher project metrics, such as the number of contributors, stars, and forks, shorter waiting times, and higher acceptance rates for pull requests. 
However, 68\% of the projects in the dataset do not currently use unit tests. We examined the projects that use unit tests by analyzing the AST trees and found that most unit tests tested the Layer unit, focusing on testing I/O.
We discuss the implications of these findings and the study's potential limitations. We suggest that developers use the taxonomies outlined in the paper as a guide for creating unit tests to improve the reliability and reproducibility of open-source DL projects. For researchers, they could explore automatic unit tests generation for DL projects or improve the existing code quality.


\section*{Acknowledgements}
This research is partially funded by the Australian Research Council under grant DP210100041.


\bibliographystyle{ACM-Reference-Format}
\bibliography{reference}

\end{document}